\documentstyle[12pt]{article}
\begin{document}
\begin{center}
\large {\bf PREDICTIONS OF THE TOP MASS IN  MINIMAL
SUPERSYMMETRIC LEFT-RIGHT MODEL} \\
\vskip 1in
Biswajoy Brahmachari \\
\end{center}

\begin{enumerate}

\item[{(a)}] Theory Group, Physical Research Laboratory, Navrangpura, \\
Ahmedabad 380 009, INDIA \\

\item[{(b)}] High Energy Section, International Centre For Theoretical
Physics,  \\ 34100 Trieste, ITALY. \\

\end{enumerate}

\noindent PACS number(s) : 11.30.Pb, 12.10 Dm
\vskip 1in
{
\begin{center}
\underbar{Abstract} \\
\end{center}

The one-loop evolution of Yukawa couplings in the minimal supersymmetric
left-right model (MSUSYLR) model with a wide variation of the
right handed breaking scale $M_R$ from 1 TeV to $10^{18}$ GeV is studied
assuming that all third generation Yukawa couplings are equal and in the
fixed point domain of the top quark Yuakwa coupling ($h_t$) at the Plank
scale. We show that: (1) The top quark Yukawa coupling
$h_t$ displays a fixed point behaviour that is similar to that of the
minimal supersymmetric standard model (MSSM). (2) The MSUSYLR model
predicts a value of the top mass in the interval 177 to 184 GeV for
$\alpha_s$ in the interval 0.11 to 0.12. (3) A large value of $tan\beta$
is required to reproduce the correct mass of the bottom quark and tau lepton.
(4) With the experimental value of the ratio ${m_b(m_b) \over m_\tau(m_\tau)}$
as an input the range of the right handed symmetry breaking scale $M_R$ can be
predicted. (5) The numerical value of the Majorana Yukawa coupling
$h_M$ can be calculated which is otherwise a completely free parameter.}
\vskip 1.5in
\begin{center}
 TO APPEAR IN PHYS REV D RAPID COMMUNICATIONS
\end{center}
\newpage
There is a lot of interest among physicists in the possible measurement of
the top quark mass $m_t$ by the CDF group in the vicinity of 174 GeV
\cite{topmass}. What does this kind of a value mean for various
theoretical models trying to generate $m_t$ ? It is well-known
that \cite{fixpt} the top quark mass $m_t$ may get fixed at an infrared
stable fixed point by the low energy structure of the renormalization
group equations (RGE) of the corresponding Yukawa coupling $h_t$. These
equations determine the evolution of $h_t$ from a large mass scale ($M_X
\simeq 10^{19} GeV $) to $m_t$. One obtains a universal value of
$h_t(m_t)$ for a large domain of values of $h_t(M_X)$. This result is
very interesting in that it shows how the details of the possibly
complicated symmetry breaking mechanisms at $M_X$ might be obliterated by
the renormalization group equations, whose fixed point structure emerges
dominant at low energies. The insensitivity to the ultraviolet behaviour
is a hallmark of infrared stable fixed points in all branches of physics.
In this communication we want to test this insensitivity by studying the
behaviour of the top Yukawa coupling in the Minimal Supersymmetric
Left Right Model (MSUSYLR) in comparison with that in the Minimal
Supersymmetric Standard Model (MSSM).

The MSSM admits an N=1 global supersymmetry by construction and
consequently the spectrum includes the superpartners of all the fermions
and bosons of the Standard Model (SM), extended to two Higgs doublets.
R-parity; $R_p=(-)^{3B+L+2S}$ ( with B,L,S as baryon number, lepton
number and spin respectively ) distinguishes between particles ($R_p=1$)
and superparticles ($R_p=-1$). The $R_p$ violating terms, if present in
the superpotential, lead to lepton and/or baryon non-conservation. Unless
one of these two conservations breakdowns is very small in magnitude,
they will induce catastrophic proton decay unobserved in nature. The
popular assumption in MSSM \cite{nilles} has been to have the
$R_p$-conservation built in by fiat though the $R_p$ violating terms are
allowed by gauge invariance and supersymmetry \cite{brahm}.

Sometime ago, it was noted that, when MSSM is extended to MSUSYLR, the
unwanted $R_p$ violating terms automatically vanish \cite{mohap}. At the
level of an underlying SO(10) GUT, this can be easily understood. SO(10)
does not allow a singlet in the product representation
$16\times16\times16$. This is a strong motivation to study the
MSUSYLR model. Of course, spontaneous $R_p$ breaking is allowed in the
MSUSYLR model, however, being spontaneous in
nature, this violation can be kept under desirable control at low energy.

Recently, a number of studies have been made of the top quark Yukawa
coupling in the Minimal Supersymmetric Standard Model (MSSM) \cite{berg}.
Here we perform a
similar study for the SUSYLR model. The respective top and bottom Yukawa
couplings $h_t$ and $h_b$ of the MSSM get embedded in the quark Yukawa
coupling $h_q$ of the MSUSYLR model at the scale $M_R$. Similarly, the tau
lepton Yukawa
coupling $h_\tau$ gets embedded in the lepton Yukawa coupling $h_l$. The
symmetry breaking chain is as shown below.

\begin{eqnarray}
&MSUSYLR&~=~SU(3)\times SU(2)_L\times SU(2)_R \times U(1)_{B-L} \nonumber\\
{M_R} \Longrightarrow~&MSSM&~=~ SU(3) \times SU(2)_L \times
U(1)_Y  \nonumber\\
{M_{SUSY}} \Longrightarrow~&SM&~=~ SU(3) \times SU(2)_L \times U(1)_Y
\nonumber\\
{M_Z} \Longrightarrow~&QED+QCD&~=~SU(3)_c\times U(1)_{em} \nonumber\\
\end{eqnarray}
There are various scenarios of gauge coupling unification in the SUSYLR
model. With the minimal choice of the Higgs fields, the right handed
symmetry breaking scale has to be comparable to the unification scale
in order to get a consistent gauge coupling unification. However, one can
enlarge the Higgs choice \cite{desh} or include the effect of the higher
dimensional operators \cite{utpal} to get a low energy right handed
symmetry breaking scale. In this paper we stick to the minimal Higgs
choice \cite{ps} and do the calculation in such a way that the results
become independent of the specific model of gauge coupling unification.
The right handed $SU(2)_R$ group is broken by the VEV of the scalar $
\overline{ \Delta_2} = (1,1,3,\sqrt{ 3 \over 2})$ under the SUSYLR
symmetry group. The field $\Delta_1=(1,3,1,\sqrt{3 \over 2})$ has to be
present to keep the left-right parity ($g_l=g_R$) intact. The scalar
field $\phi=(1,2,2,0)$ has the two MSSM Higgs doublets $H_1$ and $H_2$
embedded in it. The matter superfield representations of the MSUSYLR
model and the corresponding representation in MSSM are given in tables 1
and 2. At the right handed symmetry breaking scale, the $U(1)_Y$
hypercharge emerges as a combination of the diagonal generator of
$SU(2)_R$ and the generator of $U(1)_{B-L}$:

\begin{equation}
Y= \sqrt{3 \over 5}~T^3_R + \sqrt{2 \over 5}~{ (B-L)}
\end{equation}

The matter superfields of the MSUSYLR model embed those of the MSSM in
the following way. Thus $Q_1$ of the MSUSYLR contains the Q of the MSSM
[see table 1 and table 2] while $\overline{Q_2}$ of the MSUSYLR contains
the $\overline{U}$ and $\overline{D}$ of the MSSM. $L_1$ of the MSUSYLR
contains L of the MSSM while $\overline{L_2}$ of the contains
$\overline{E}$. The right handed neutrino gets a large Majorana mass at
the right handed symmetry breaking scale.

\begin{table}[htb]
\begin{center}
\[
\begin{array}{|c||c||c|}
\hline
Superfield & SU(3)_c \times SU(2)_L \times SU(2)_R \times U(1)_{B-L}&
Anomalous~ Dimension \\
\hline
Q_1&(3,2,1,{1 \over 6} \sqrt{3 \over 2}) & {1 \over 16~\pi^2} [ 2h^2_Q -
{8 \over 3} g^2_c-{3 \over 2} g^2_L - { 1 \over 12} g^2_{B-L}] \\
\overline{Q_2} & (\bar{3},1,2,-{ 1\over 6} \sqrt{ 3 \over 2})& { 1 \over 16
\pi^2} [2 h^2_Q - { 8 \over 3} g^2_c - { 3 \over 2} g^2_R - { 1\over 12 }
g^2_{B-L}] \\
L_1 & (1,2,1,- {1\over 2}\sqrt{ 3 \over 2}) & { 1\over 16 \pi^2} [2 h^2_L
+ 2 h^2_M - { 3 \over 2} g^2_L - { 3 \over 4} g^2_{B-L} ] \\
\overline{L_2} & (1,1,2, {1 \over 2}\sqrt{ 3 \over 2}) & { 1 \over 16 \pi
^2} [ 2 h^2_L + 2 h^2_M - { 3 \over 2} g ^2_R - { 3 \over 4} g ^2 _{B-L} ]\\
\Delta_1 & ( 1,3,1,\sqrt{ 3 \over 2} ) & { 1 \over 16 \pi^2}[ h^2_M - 4
g^2_L - 3 g^2_{B-L}] \\
\overline{ \Delta_2} & (1,1,3,-\sqrt{3 \over 2}) & { 1 \over 16 \pi^2} [
h^2_M - 4 g^2_R - 3 g^2_{B-L}] \\
\phi & ( 1,2,2,0) & { 1 \over 16 \pi^2} [ 3 h^2_Q + h^2_l - { 3 \over 2}
g^2_L - { 3 \over 2} g^2_R] \\
\hline
\end{array}
\]
\end{center}
\caption{The Superfields in the MSUSYLR model. Representations and the
anomalous dimensions }
\label{table1}
\end{table}

The Lagrangian density of MSSM in standard superfield notation is given
by,
\begin{equation}
{\cal L}=
h_\tau~[L~H_1~\overline{E}]_F+~h_b~[Q~H_1~\overline{D}]_F
+h_t~[Q~H_2~\overline{U}]_F,
\end{equation}
while that of the MSUSYLR is given by \cite{kuchi}
\begin{equation}
{\cal L}=h_Q~[Q^T_1~\tau_2 \phi \tau_2~\overline{Q}_2]_F + h_l~
[L^T_1~\tau_2\phi\tau_2~\overline{L}_2]_F+ ih_M~[L^T_1~\tau_2\Delta_1 L_1+
\overline{L}^T_2~\tau_2 \overline{\Delta}_2~\overline{L}_2]_F
\end{equation}
\begin{table}[htb]
\begin{center}
\[
\begin{array}{|c||c||c|}
\hline
Superfields& SU(3)_c \times SU(2)_L \times U(1)_Y & Anomalous~Dimension\\
\hline
L& (1,2,-{1 \over 2} \sqrt{ 3 \over 5}) & { 1 \over 16 \pi^2}~[ h^2_\tau
- { 3 \over 2} g^2_2 - { 3 \over 10} g^2_Y ]\\
\overline{E}&(1,1,\sqrt{ 3 \over 5}) & { 1 \over 16 \pi^2 }~[2 h^2_\tau
- { 6 \over 5} g^2_Y]  \\
\overline{D}& (\overline{3},1,{ 1 \over 3} \sqrt{ 3 \over 5}) & { 1 \over
16 \pi^2}~[2 h^2_b-{ 8 \over 3} g^2_c - { 4 \over 30} g^2_Y] \\
\overline{U}&(\overline{3},1,-{2 \over 3} \sqrt{ 3 \over 5})& { 1 \over 16
\pi^2}~[ 2 h^2_t-{ 8 \over 3} g^2_3 - { 8 \over 15} g^2_Y] \\
Q&(3,2,{ 1 \over 6} \sqrt{ 3 \over 5}) & { 1 \over 16 \pi^2}~[ h^2_t +
h^2_b - { 8 \over 3} g^2_3 - { 3 \over 2} g ^2_2 - { 1 \over 30} g^2_Y]\\
H_1&(1,2,-{ 1 \over 2} \sqrt{ 3 \over 5})& { 1 \over 16 \pi^2}~[h^2_\tau+
3 h^2_b - { 3 \over 2} g^2_2 - { 3 \over 10} g^2_Y]\\
H_2&(1,2,{ 1 \over 2} \sqrt{ 3 \over 5}) & { 1 \over 16 \pi^2}~[3 h^2_t - {
3 \over 2} g^2_2 - { 3 \over 10} g^2_Y]\\
\hline
\end{array}
\]
\end{center}
\caption{Superfields in MSSM: Representations and anomalous dimensions }
\label{table2}
\end{table}
Renormalization group  equations constitute our basic tool in studying the
evolution of
the  relevant Yukawa coupling. Given a trilinear term $d_{abc}~\Phi^a
\Phi^b\Phi^c$ in
the superpotential and the evolution scale $\mu$, the RGE for $d_{abc}$
is \cite{west}
\begin{equation}
\mu { \partial  \over  \partial \mu} d_{abc} = \gamma^i_a d_{ibc} +
\gamma^j_b  d_{ajc} + \gamma^k_c d_{abk}  \label{yrg}
\end{equation}
In  Eqn. \ref{yrg},~$\gamma^i_a$  is the anomalous dimension matrix
\begin{equation}
\gamma^i_a={Z^{-1/2}_a}^k~{\mu {\partial \over \partial
\mu}}~{Z^{1/2}_k}^i = {1\over {16 \pi^2}} ~[n_p d^2 - 2
\delta^i_a \sum_k g^2_k c^k_A]\end{equation}
Where, Z is the renormalization constant matrix relating  the renormalized
superfield
$\Phi$ to the unrenormalized superfield   $\Phi_0$ by the relation,
\begin{equation}
\Phi^i_0 =  {Z^{1/2}_a}^i~\Phi^a
\end{equation}

$n_p$ is a numerical factors denoting the number of possible graphs and
in usual notation,
\[
[\sum_i T^i T^i]_{mn}=c_A~\delta_{mn}.
\]
We shall apply Eqn. \ref{yrg} to the
Yukawa couplings of interest.  First  we use this
equation and the informations tabulated in Table. \ref{table1}
and  Table. \ref{table2}
to get the evolution equations of the Yukawa  couplings  in the
MSSM and the MSUSYLR
model. Defining $\alpha_i={g^2_i \over 4 \pi}$ and  $Y_i={h^2_i
\over 4 \pi}$, we can write those equations. In the region above $M_R$ we
have;
\begin{eqnarray}
{\partial Y_Q \over \partial t}
&=&[7Y_Q+Y_l-{16 \over 3} \alpha_3-3 \alpha_{2L}-3 \alpha_{2R} -{1
\over 6} \alpha_{B-L}] Y_Q, \nonumber\\
{\partial Y_l \over \partial t}
&=&[3 Y_Q + 5Y_l+4 Y_M-3 \alpha_{2L} - 3 \alpha_{2R}- { 3 \over 2}
\alpha_{B-L}] Y_l, \nonumber\\
{\partial Y_M \over \partial t}
&=&[4 Y_l + 5 Y_M - 7 \alpha_{2L}-{ 9 \over 2} \alpha_{B-L}] Y_M,
\nonumber\\
{\partial \alpha_3 \over \partial t}&=& [-9+6] ~\alpha^2_3, \nonumber\\
{\partial \alpha_{2L} \over \partial t}&=& [-6+6+3] ~\alpha^2_{2L},
\nonumber\\
{\partial \alpha_{2R} \over \partial t}
&=& [-6+6+3] ~\alpha^2_{2R}, \nonumber\\
{\partial \alpha_{B-L} \over \partial t}&=& [-0+6+9]
{}~\alpha^2_{B-L}, \label{rglr}
\end{eqnarray}
while, in the region $ M_{SUSY} < \mu < M_R$ we have,

\begin{eqnarray}
{\partial Y_t \over \partial t}
&=& [6 Y_t+Y_b-{16 \over 3} \alpha_3 -3\alpha_2-{ 13 \over 15} \alpha_Y]
Y_t, \nonumber\\ {\partial Y_b \over \partial t}
&=& [6 Y_b+ Y_t+Y_\tau - { 16 \over 3} \alpha_3 - 3 \alpha_2 - { 7 \over
15} \alpha_Y] Y_b, \nonumber\\
{ \partial Y_\tau \over \partial t}
&=& [4 Y_\tau + 3 Y_b - 3 \alpha_2 - { 9 \over 5} \alpha_Y]
Y_\tau, \nonumber\\
{\partial \alpha_3 \over \partial t}&=& [-9+6]~\alpha^2_3, \nonumber\\
{\partial \alpha_{2l}\over \partial t} &=& [-6+6+1]~\alpha^2_{2L},
\nonumber\\
{\partial \alpha_{1Y} \over \partial t}&=& [-0+6+
{3 \over 5}] ~\alpha^2_{1Y}.
\label{rgsm}
\end{eqnarray}
In Eqn. \ref{rglr} and Eqn. \ref{rgsm} we have used $t={1 \over {2 \pi}}
{}~( ln {\mu \over 1 GeV})$. For the gauge coupling evolutions the 1 loop beta
functions are very well known. For completeness we give the generic
formula for the beta function. In our notation we have,
\[
\beta=[-3~N + 2~n_f + T_s]
\]
where, the first term comes form the gauge contribution and  the second
term comes from the fermionic contribution. The variable N refers to the
gauge group SU(N) and $n_f$ signifies the number of fermionic generations.
The last term, $T_S$, represents the contribution of the scalars.

In our calculation we have assumed $M_{SUSY}= 1 TeV$
as such a scale may solve the natural ness problem in
the Higgs sector. From the electroweak scale $M_Z$ to the supersymmetry
breaking scale $M_{SUSY}$ the evolution of the couplings are governed
by non-supersymmetric renormalization group equations {\cite {mv}}

In brief, we have adopted the following procedure. We assume that MSUSYLR
symmetry holds good up to a large cut-off scale $M_X=10^{19}$ GeV. We do
not require the gauge couplings to unify. On the other hand, all third
generation Yukawa couplings of the MSUSYLR model have been taken to be of
order one  at $M_X$ as exhibited in Eqn. \ref{inputs}. This scenario of
Yukawa couplings in the ultraviolet region is a
predictive one, as shown by Hill \cite{fixpt}, in the sense that
$h_t(m_t)$ is insensitive to the variation of  $h_t(M_X)$ within this
region. In a more mathematical terminology, $h_t(M_X)$ stays in the domain of
attraction. We also think that such an assumption is justified as
the third generation fermions are much heavier compared to the rest.
The renormalization group equations are solved numerically with the inputs;

\begin{equation}
{h^2_i(M_X) \over 4 \pi}=1,~
\alpha_3(M_Z)=0.11-0.12,~\alpha_{2L}(M_Z)=0.03322,~\alpha_{1Y}(M_Z)=0.01688.
\label{inputs}
\end{equation}

We see that as $\mu$ decreases the couplings evolve downwards very fast
and reach a fixed point at the infrared region. We have varied the right
handed symmetry breaking scale $M_R$ in a very wide range; from the TeV
region  to $10^{18}$ GeV. The value of the
top quark, the bottom quark and the tau lepton Yukawa couplings have been
calculated at
the scale $M_t=170$ GeV [$h_t(M_t)$, $h_b(M_t)$ and $h_\tau(M_t)$] with
respect to this wide range of
variation in the right handed symmetry breaking scale. These results are
summarized in the Table \ref{table3} and Table \ref{table4}. The value of
the Majorana Yukawa
coupling at the right handed symmetry breaking scale [$h_M(M_R)$] also
emerges from this analysis. These values are also tabulated in
Table \ref{table3} and Table \ref{table4}.
Note that below the scale $M_R$ we have considered only the MSSM
couplings whereas in general one has a light left handed
triplet below $M_R$. This triplet, when present below $M_R$, will not have
any tree level coupling with the  top quark which ensures that our results
will be very nearly valid even in that case. The scenario with a low energy
triplet and its phenomenological consequences will be studied elsewhere
\cite{else}.

\begin{table}[htb]
\begin{center}
\[
\begin{array}{|c||c||c||c||c||c||c||c||c|}
\hline
M_R&h_t(M_t)&h_b(M_t)&h_\tau(M_t)& h_M(M_R)&
tan\beta&m_b(M_t)&m_t(M_t)&{m_b(M_t) \over m_\tau(M_t)}\\
\hline
10^3   &1.02&1.02&0.33&1.16&32.71&5.45&178.7&2.71  \\
10^4   &1.01&1.01&0.34&1.17&33.67&5.22&176.7&2.61  \\
10^6   &0.99&0.98&0.36&1.20&35.84&4.80&173.7&2.43  \\
10^8   &0.98&0.97&0.39&1.24&38.30&4.42&171.8&2.29  \\
10^{10}&0.98&0.96&0.42&1.30&41.13&4.06&171.6&2.11  \\
10^{12}&0.97&0.95&0.45&1.39&44.41&3.72&170.0&1.96  \\
10^{14}&0.97&0.94&0.49&1.55&48.33&3.39&169.8&1.87  \\
10^{16}&0.97&0.93&0.54&1.90&53.25&3.04&170.1&1.65  \\
\hline
\end{array}
\]
\end{center}
\caption{The values of $h_t(M_t)$, $h_b(M_t)$, $h_\tau(M_t)$ and
$h_M(M_R)$ calculated by RGE up to the second decimal place for
$\alpha_s=0.11$. The prediction of the
masses $m_b$ and $m_t$ defined by the Eqn. 12 and Eqn. 13 at the scale
$M_t$ has been quoted in GeV. $M_t$ is defined as 170 GeV.}
\label{table3}
\end{table}

\begin{table}[htb]
\begin{center}
\[
\begin{array}{|c||c||c||c||c||c||c||c||c|}
\hline
M_R&h_t(M_t)&h_b(M_t)&h_\tau(M_t)& h_M(M_R)&
tan\beta&m_b(M_t)&m_t(M_t)&{m_b(M_t) \over m_\tau(M_t)}\\
\hline
10^3   &1.05&1.04&0.33&1.16&32.33&5.63&182.6&3.17  \\
10^4   &1.03&1.03&0.34&1.17&33.31&5.39&180.5&3.03  \\
10^6   &1.02&1.01&0.36&1.20&35.46&4.96&177.5&2.79  \\
10^8   &1.00&0.99&0.38&1.24&37.91&4.56&175.6&2.57  \\
10^{10}&1.00&0.98&0.41&1.30&40.71&4.20&174.4&2.36  \\
10^{12}&1.00&0.97&0.44&1.39&43.98&3.85&173.8&2.16  \\
10^{14}&0.99&0.96&0.48&1.55&47.87&3.50&173.7&1.97  \\
10^{16}&0.99&0.95&0.53&1.90&52.77&3.15&173.9&1.77  \\
10^{17}&1.00&0.95&0.57&2.36&55.91&2.95&174.2&1.66 \\
\hline
\end{array}
\]
\end{center}
\caption{The values of $h_t(M_t)$, $h_b(M_t)$, $h_\tau(M_t)$ and
$h_M(M_R)$ calculated by RGE up to the second decimal place for
$\alpha_s=0.12$. The prediction of the
masses $m_b$ and $m_t$ defined by the Eqn. 12 and Eqn. 13 at the scale $M_t$
has been quoted in GeV. $M_t$ is defined as 170 GeV.}
\label{table4}
\end{table}

The value of $tan\beta$, defined as ${<H_2> \over
<H_1>}$, can be estimated from the measured  value \cite{pdg} of the tau
lepton mass, having little experimental error, by the equation,
\begin{equation}
m_\tau(m_\tau) \simeq m_\tau(M_t)= h_\tau(M_t)~174~cos\beta=1.777 ~GeV,
\end{equation}
where, $H_2$ and $H_1$ are the two Higgs doublets embedded in the Higgs
$\phi$ of MSUSYLR.
The estimated value of $tan\beta$ is tabulated in the sixth columns of
Table \ref{table3} and Table \ref{table4}. Once the value of $tan\beta$ is
known the
predictions for the top mass and the bottom \footnote{ I thank referee
for important comments on the prediction of $m_b$.} mass follows
from the equations,

\begin{eqnarray}
m_b(M_t)&=&h_b(M_t)~174~cos\beta, \label{mb}\\
m_t(M_t)&=&h_t(M_t)~174~sin\beta, \label{mt}
\end{eqnarray}

which are tabulated in the seventh and eighth columns of Table \ref{table3}.
The prediction for the pole mass of the top quark immediately follows
from the equation,

\begin{equation}
m_t(~pole~) = m_t(M_t)~[1+{4 \over 3 \pi}\alpha_s(M_t)+O(\alpha^2_3)].
\label{pole}
\end{equation}

At this stage we consider the last column of Table \ref{table3} and Table
\ref{table4}.
The ratio can be estimated from the experimental numbers by the relation,

\begin{equation}
{m_b(M_t) \over m_\tau(M_t)}={m_b(m_b) \over m_\tau(m_\tau)} {\eta_\tau
\over \eta_b},
\end{equation}

where $\eta_\tau$ and $\eta_b$ parametrize \cite{berg} the evolution of the
masses from their respective scales to the scale $M_t$. We take the value of
${\eta_\tau \over \eta_b}$ to be 0.74 for $\alpha_s$=0.11 and 0.67 for
$\alpha_s$=0.12. Allowing
the value of $m_b(m_b)$ in the interval of 4.1 GeV to 4.5 GeV \cite{pdg}
we get the range,

\begin{eqnarray}
1.6<{m_b(M_t) \over m_\tau(M_t)}<1.9&for&\alpha_s=0.11-0.12.
\end{eqnarray}

Now, we can read off a bound on the mass scale $M_R$ from the ninth
columns of Table \ref{table3} and Table \ref{table4},

\begin{eqnarray}
10^{12} ~GeV < M_R <10^{17}~GeV &for& \alpha_s=0.11-0.12. \label{bound1}
\end{eqnarray}

We have taken a conservative lower bound on $M_R$
realizing that $m_b(m_b)$ may be somewhat larger than 4.5 GeV as well
\cite{nub}. Now, combining the results in the eighth columns
of Table \ref{table3} and Table \ref{table4} and in Eqn. \ref{bound1}
together with Eqn. \ref{pole} we get the one-loop
prediction of the physical top mass.

\begin{eqnarray}
177.9 < m_t(~pole~) < 183.2 ~GeV  &for& \alpha_s=0.11-0.12 \label{bound2}
\end{eqnarray}

It is interesting that this scenario of SUSYLR
model can predict the top mass in the expected range \cite{topmass} and
also satisfy the measured values of $m_b(m_b)$ and $m_\tau(m_\tau)$.

The lower bound of $M_R$ displayed in Eqn. \ref{bound1}
calls for some comments. There are hints of such a large value
of $M_R$ from gauge coupling unification. Our study shows that even
independent of the unification of gauge couplings, to predict the correct
range of values of the low energy ratio of $m_b \over m_\tau$, we need a
large value for the scale $M_R$. We consider it to be a welcome result as
a confirmation of the popular MSW \cite{msw} solution of the solar
neutrino problem will also suggest that $M_R$ is in the range
$10^{10}-10^{12}$ GeV.

To conclude, in this paper we have solved the
renormalization group equations of the Yukawa couplings in the SUSYLR model
numerically. We have :

\noindent (1) Shown that the fixed point value of the top Yukawa coupling is
insensitive to the variation of the right handed symmetry breaking scale.
Bound obtained on the top mass is given in Eqn. \ref{bound2}.

\noindent (2) Shown that the low energy ratio of $m_b \over m_\tau$ is
sensitive to the right handed symmetry breaking scale,
which can be used to predict that $10^{12} <M_R <10^{17}$ GeV.

\noindent (3) Calculated the numerical value of the Majorana Yukawa
coupling $h_M$ at the right handed symmetry breaking scale; this coupling
is otherwise a free parameter of the SUSYLR model.

I would like to acknowledge the hospitality of the Tata Institute
of Fundamental Research, Bombay,  where a part of this work was done.
This work was initiated at WHEPP3 meeting held at Madras, under the
financial sponsorship of S. N. Bose National Centre For Basic
Sciences, Calcutta, and Institute of Mathematical Sciences, Madras.
I also thank K. S. Babu, Probir Roy and A. Y. Smirnov for
discussions and R. N. Mohapatra for private communications.

\end{document}